\documentclass[a4paper,10pt]{article}
\usepackage{authblk}
\usepackage{vmargin}
\usepackage{epstopdf}
\usepackage[utf8]{inputenc}
\usepackage{stackrel}
\usepackage{amsmath}

\setpapersize{A4}
\setmargins{1.5cm}       
{1cm}                        
{17.5cm}                      
{25cm}                    
{10pt}                           
{1cm}                           
{0pt}                             
{1cm}

\usepackage{graphicx}
\usepackage[superscript,biblabel]{cite}
\usepackage[utf8]{inputenc} 
\title{Reinforcement Learning of Artificial Microswimmers}

\author[1]{Santiago Mui\~{n}os-Landin}
\author[2,3]{Keyan Ghazi-Zahedi}
\author[1]{Frank Cichos}	

\affil[1]{Molecular Nanophotonics, Peter Debye Institute for Soft Matter Physics, Leipzig, Germany}
\affil[2]{Information Theory of Cognitive Systems, Max Planck Institute for Mathematics in the Sciences, Leipzig, Germany}
\affil[3]{Santa Fe Institute, Santa Fe, NM, USA}
\date{\today}

\begin{document}
\maketitle

\textbf{The behavior of living systems is based on the experience they gain through their interactions with the environment \cite{Arnott2017}. This experience is stored in complex biochemical networks of cells and organisms to provide a relationship between a sensed situation and what to do in this situation\cite{Pomerol1997,Shadlen2009,Dayan2006}. Inspired by this form of adaption machine learning algorithms have been developed and implemented  \cite{Kober2012,Nagpal2014}. However, for an artificial microswimmer, a microparticle that mimics the propulsion as one of the basic functionalities of living systems \cite{Gompper2016, Bechinger:2016cf} such adaptive behavior and learning has not been demonstrated so far.
Here we introduce machine learning algorithms to the field of artificial microswimmers  with a hybrid approach. We employ self-thermophoretic active particles\cite{Jiang2010,Qian2013} in a real world environment  which are controlled by a real-time microscopy system to introduce reinforcement learning\cite{Littman2015,Kaelbling96,Sutton1998}. We demonstrate the solution of a standard problem of reinforcement learning -- the navigation in a gridworld. Due to the size of the microswimmer, noise introduced by Brownian motion is found to contribute considerably to both the learning process and the actions within a learned behavior. We extend the learning process to multiple swimmers and sharing of information during the learning process. Our work represents a first step towards the integration of learning strategies into microsystems and provides an experimental platform for the study of the emergence of adaptive and collective behavior.}


\begin{figure}[!htb]
	\centering
	\includegraphics[width=1\columnwidth]{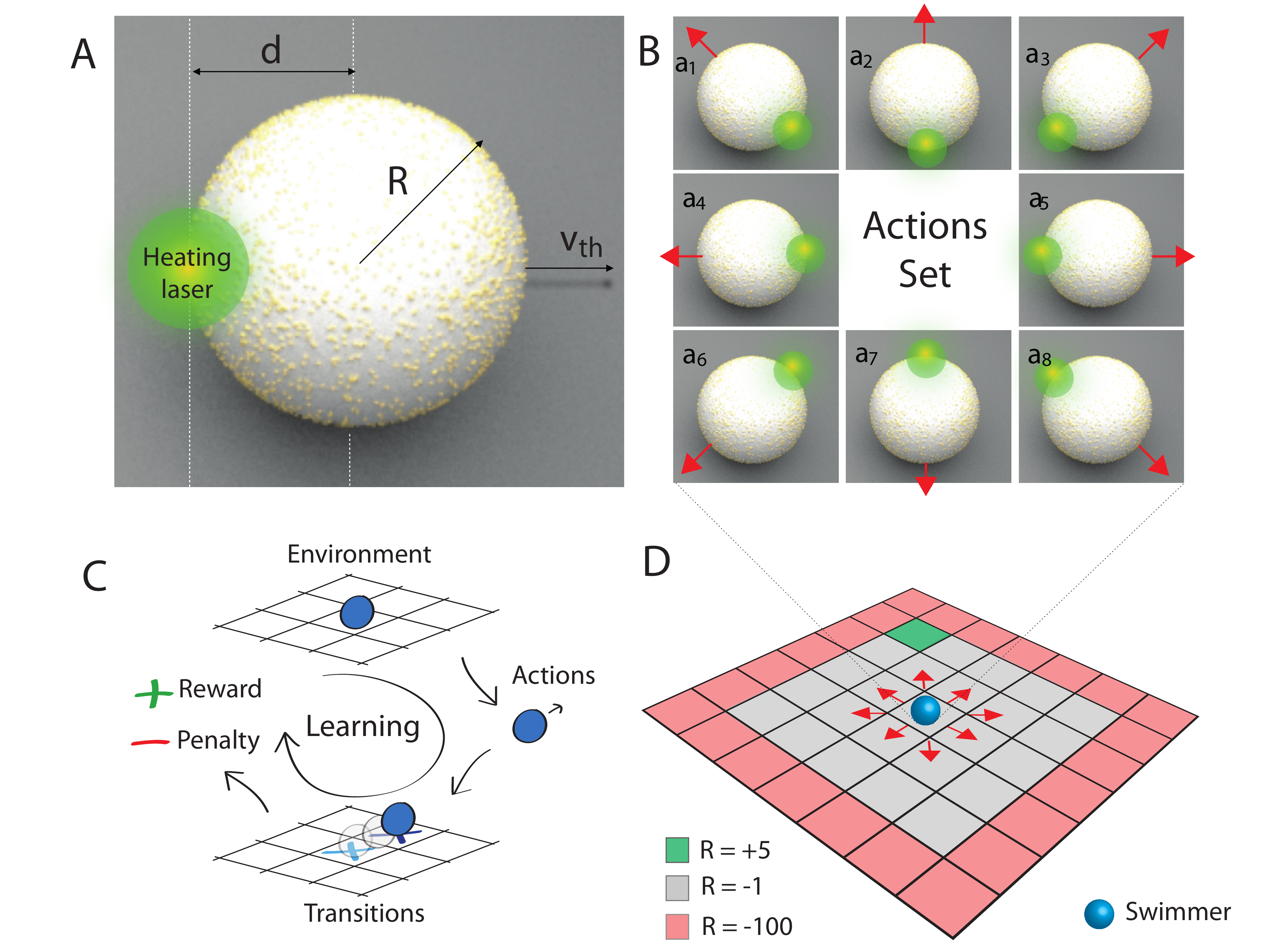}
	\caption{{\bf Swimmer, swimmer states and actions.} {\bf A)} Sketch of the self-thermophoretic symmetric microswimmer. The used particles have an average radius of $r = 1.09$ $\rm \mu m$ and are covered on 30 \% of their surface with gold nanoparticles of about 10 nm diameter. A heating laser illuminates the colloid asymmetrically (at a distance $d$ from the center) and the swimmer acquires a well defined thermophoretic velocity $v$. {\bf B)} Different actions are implemented by setting a specific heating position on the circumference of the particle, which defines the propulsion direction. Overall 8 different actions are used in the experiments. {\bf C)} Concept of the applied reinforcement learning algorithm consisting of the detection of the state (position) of the swimmer in the environment, an action changing the state of the swimmer and awarding corresponding rewards for the transition. {\bf D)} Grid of states (gridworld) defined for the experiments where each cell represents a state. Through the execution of different actions (see B) the swimmer is carrying out transitions between the different states collecting rewards (R) that can be positive or negative (penalty) as indicated in the legend.}
	\label{fig1}
\end{figure}	

Living organisms adapt their behavior according to their environment to achieve a particular goal. Information about the state of the environment is sensed, processed and encoded in the organism to provide the appropriate future actions or properties in return. Such learning or adaptive processes occur within the lifetime of a generation, over multiple generations or over evolutionary relevant timescales.  Swarms of fish or flocks of birds have developed a collective behavior which adapts to the existence of predators \cite{parrish_hamner_1997} and collective hunting may represent a more efficient foraging strategy \cite{Abaid2013}. Birds learn how to use convective air flows \cite{Reddy:2016ef}. Sperms have evolved complex swimming patterns to explore chemical gradients in chemotaxis\cite{Friedrich2015} or bacteria express specific shapes to follow gravity\cite{Bechinger2014}. 
Inspired by these adaptive and optimization processes, learning strategies have been developed \cite{Kaelbling96}, which reduce the complexity of the physical and chemical processes to a  mathematical procedure. Many of these learning strategies have been implemented into robotic systems \cite{Kober2012}. One particular framework is reinforcement learning (RL) in which a so-called agent gains experience by interacting with its environment. The value of his experience relates to rewards (or penalties) connected to the states the agent can occupy. Such reinforcement signals can be given at each state or only very sparsely in a single target state \cite{Dayan2006}. The learning process then maximizes the cumulative reward for a chain of actions to obtain the so called {\it policy}. This policy then finally advices the agent which action to take to achieve the maximum cumulative reward. Recent computational studies, for example, reveal that reinforcement learning can provide optimal strategies for the navigation through flows \cite{Biferale2017}, the swarming of robots \cite{La2015}, the soaring of birds \cite{Reddy:2016ef} or the development of collective motion \cite{Ried:it}.
All of the afore mentioned examples require the agent to be able to carry out an action e.g. in form of a directed motion. Artificial microswimmers have recently gained considerable interest as a class of active material as they provide this basic biological functionality of directed motion by self-propulsion in an artificial system \cite{Bechinger:2016cf}. The non-equilibrium character of these systems already reveal new collective phenomena \cite{Palacci:2013eu, Buttinoni:2013de}. An incorporation of adaptive behavior would extend this, but has to our knowledge not been achieved so far. While the integration of the necessary components into an artificial microswimmer for a self-learning still requires further technological progress, external control schemes for swimmers may provide an equivalent result.
In this report we incorporate algorithms of reinforcement learning into the motion of artificial microswimmer in an aqueous solution with the help of an optical control scheme \cite{Qian2013, Bregulla2014}. The microswimmer learns to navigate in a gridworld, which is a standard problem of RL. It masters to move towards a single target, to avoid obstacles and to share the gained information with other particles in the gridworld. As micrometer-sized active particles are subject to Brownian motion we are able to explore the influence of noise on the learning process. Despite the fact that the outcome of the learning process is conceivable, our demonstration of RL for single and multiple swimmers provides the starting point to interweave the realms of machine learning and active materials.
 
\begin{figure*}[!htb]
	\centering
	\includegraphics[width=1\textwidth]{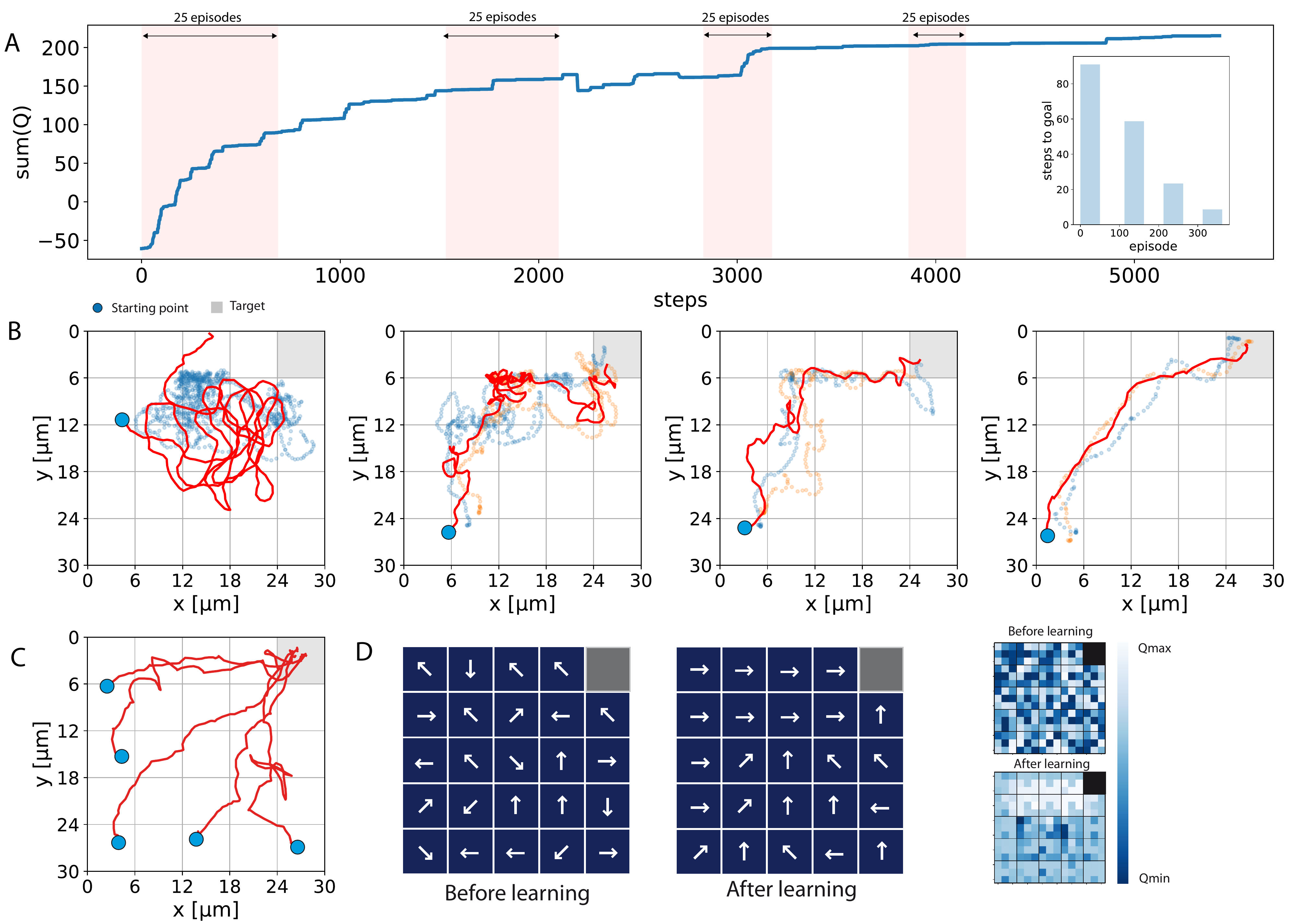}
	\caption{{\bf Single Microswimmer Learning. }{\bf A)} Evolution of the sum of all elements of the Q-matrix at each step of the learning process. The matrix is randomly  initialized. A learning step is carried out when the swimmer makes a transition between two states. The Q-matrix sum converges due to the learning process with increasing number of steps. The shaded regions denote a set of 25 episodes in the learning process, where the starting point is randomly chosen. The inset displays the average number of steps towards the goal as a function of the episode number when starting at the lower left corner.
		{\bf B)} Different examples of the behavior of a single microswimmer at different stages of the learning process. The first example corresponds to a swimmer starting at the beginning of the learning process at an arbitrary position in the gridworld. The trajectory is characterized by a large number of loops. With an increasing number of learning episodes, the trajectories become more persistent in its motion towards the goal. This is also reflected by the decreasing average number of steps taken to reach the goal (see inset in A).  {\bf C)} Example trajectories for the swimmer starting at different states after a nearly converged Q-matrix is obtained. {\bf D)} Policies defined by the Q-matrix before and after the convergence of the learning process. On the right is a color representation of the Q-matrix for the corresponding policies. The figure shows measurements done at Pe = 80. } 
	\label{fig2}		
\end{figure*}	

Our microswimmer -- the agent in the RL language -- consists of a melamine resin particle (2.19 $\mathrm{\mu m}$ diameter) covered homogeneously on $30\,\%$ of its surface with gold nanoparticles of 10 $\rm nm$ diameter  (Fig. \ref{fig1}A). To self-propel, the swimmer has to break the time symmetry of low Reynolds number hydrodynamics \cite{Purcell77}. This symmetry breaking is achieved by an asymmetric illumination of the particle with laser light of 532 $\rm nm$ wavelength. The laser light is absorbed by the gold nanoparticles and generates a temperature gradient along the surface of the particle, which induces thermo-osmotic surface flows and finally results in a self-propulsion of the whole particle. The direction of propulsion is set by the vector pointing from the laser position to the center of the particle. To maintain the asymmetric illumination during its motion we follow the swimmers position in real time and steer the heating laser correspondingly using an acousto-optic deflector (see methods). As compared to other types of swimmers \cite{Qian2013, Jiang2010, Bechinger2014}, this symmetric swimmer removes the timescale of rotational diffusion from the swimmers motion and provides an enhanced steering accuracy \cite{Selmke:2017gya, Selmke:2017gy}.

To learn to navigate in the sample, we coarse grain the sample region of 30 $\rm \mu m$ x 30 $\rm \mu m$ into a gridworld of 25 states ($s$, 5$\times$5), each state having a dimension of 6 $ \rm \mu m $ $\times$ 6 $\rm \mu m$ (Fig. \ref{fig1}D). One of the states is defined as the target state (goal) which the swimmer is learning to reach. The obtained real-time swimmer position is used to identify the state in which the swimmer currently resides. To move between states, we define 8 actions $a_i$ ($i = 1,..,8$). The actions are carried out by placing the heating laser at the corresponding position on the circumference of the particle (see Fig. \ref{fig1}B).  A sequence of actions defines a path in the gridworld which either ends when the swimmer leaves the gridworld or reaches the target state. This defines a learning episode. To reinforce a certain behavior of the swimmer, rewards or penalties are given. In particular, the microswimmer gets a reward once it reaches the target state and a penalty in other cases (see SI for details on the reward definitions). The reward function $R$ is thus only depending on the state $s$.

For our experiments we have chosen the Q-learning \cite{Watkins1989} as the algorithm for finding an optimal policy for the navigation problem. In this framework the gained experience of the agent is stored in the Q-matrix\cite{Sutton1998} which tracks the utility of the different {\it actions} $a$ in each {\it state} $s$. When the swimmer changes between the states $s$ and $s^{\prime}$ the utility matrix is updated according to 

\begin{align}
Q_{t+\Delta t}(s,a) & = Q_t(s,a) +
\alpha\big[R(s)\nonumber\\ &\phantom{=} + \gamma
\max_{a'}Q_t(s',a')-Q_t(s,a)\big]\label{eq1}
\end{align}

The Q-matrix is changed by taking into account the reward for the current state $R(s)$, the utility of the next state $Q_t(s',a')$ after taking the best action $a'$ and the current utility $Q_t(s,a)$. The influence of these values is controlled by two factors, the learning rate $\alpha$ and the discount factor $\gamma$. The learning rate defines the speed at which new information is incorporated into the Q-matrix and the discount factor determines the current value of future events in the learning process. The reward function is the only feedback signal that the system receives to figure out what it should learn. 
The result of this RL procedure is then the policy function $\pi(s) \rightarrow a$, which is obtained by analyzing the Q-matrix, which represents the current information of the system $\pi(s)  = \max_{a} Q(s,a)$. 

As compared to purely deterministic systems, the current microswimmer in a gridworld deserves some additional comments. All experiments are carried out with a sampling timescale, which is the exposure time $\Delta t_{\rm exp}$ of the camera and equivalent to the inverse framerate in our experiments. When applying an action the system takes several exposure times depending on swimmer velocity and grid size until it transits to a new state and the Q-matrix is updated. A swimmer velocity of $v=2$  $\rm \mu m/s$ leads to a displacement of 6 $\rm \mu m$ within 3 $\rm s$, which corresponds to about 17 frames at an inverse framerate of $\Delta t_{\rm exp}=$180 $ {\rm ms}$. Thus, one transition between states (denoted as one step) in the gridworld consists out of several 10 imaging frames. A learning episode, i.e. the sequence of actions until the swimmer reaches the boundary or the goal, consists of multiple steps depending on the learning progress. Further, the microswimmer position itself is subject to noise as it carries out Brownian motion with a diffusion coefficient of $D=0.23$  $\rm \mu m^2/s$. The state change targeted by the applied action is therefore not necessarily the state change which finally occurs. For example, if the action to move horizontally to the right is applied, the swimmer might also end up in the state to the lower right as Brownian motion is driving the swimmer in the vertical direction downwards.
The level of stochasticity in the motion is quantified by the Peclet number $Pe=l v /D$. For a grid element size $l=6$ $\rm \mu m$, a velocity of $v=2$  $\rm \mu m/s$ and a diffusion coefficient of $D=0.23$ $\rm \mu m^2/s$ the Peclet number is $Pe=52$ stating that the deterministic motion is stronger than the diffusive one. Yet Brownian motion may influence the learning process in speed and the final {\it policy}. 

	\begin{figure*}[!htb]
	\centering
	\includegraphics[width=1\textwidth]{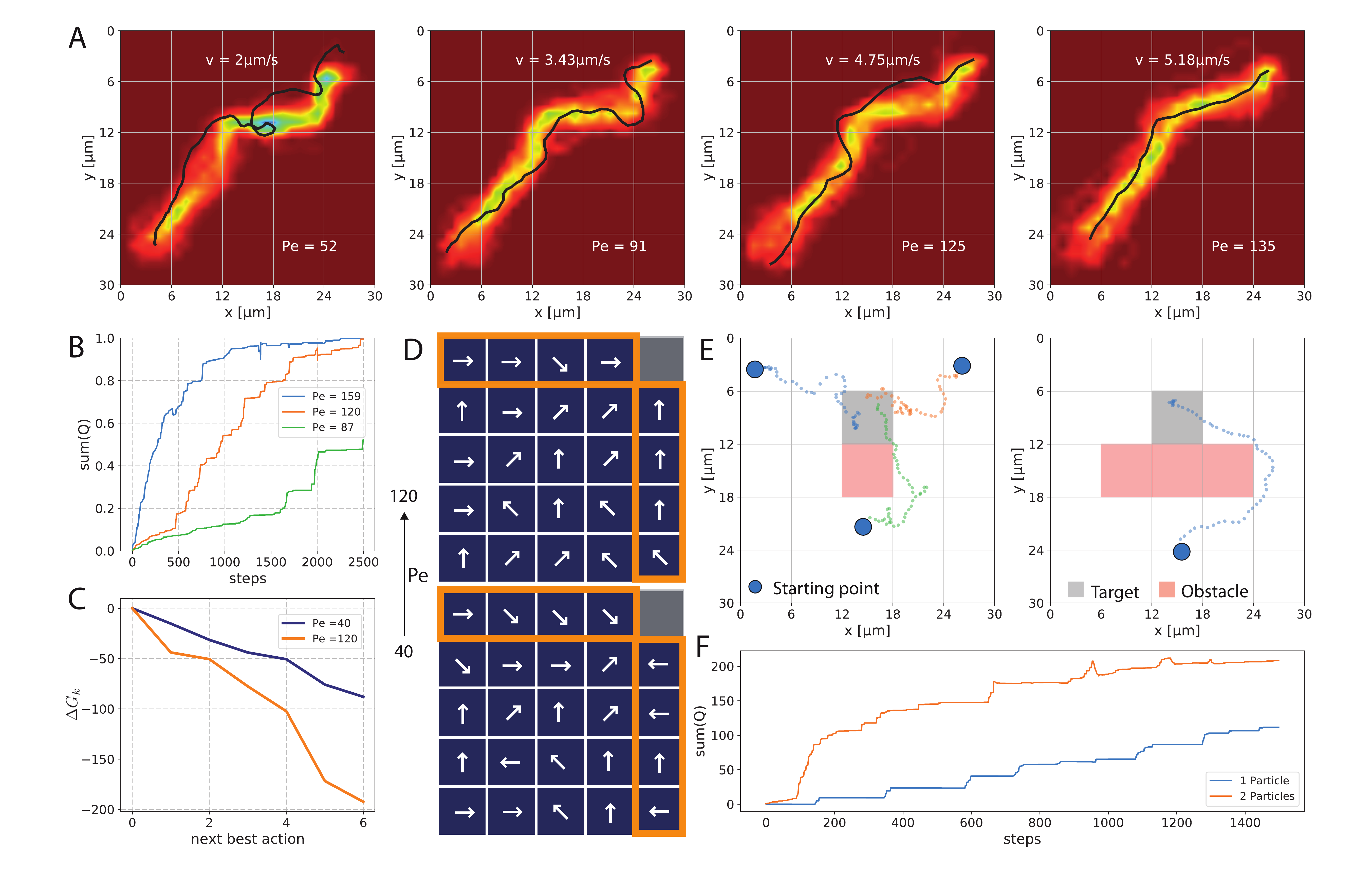}
	\caption{ {\bf Influence of Brownian Motion on the Learning Process.} {\bf A)} Swimmer position density maps from 50 trajectories of a single swimmer starting in the lower left corner of the gridworld and moving at different speeds ($2\, {\rm \mu m/s}$, $3.43\, {\rm \mu m/s}$, $4.75\, {\rm \mu m/s}$ and $5.18\, {\rm \mu m/s}$). The goal is located at the top right corner. The black line shows an example trajectory. All trajectories are recorded after a policy has been obtained from a learning process with a Peclet number of $Pe=80$. {\bf B)} Peclet number dependence of the learning progress as defined by the sum of all Q-matrix elements normalized to the value at convergence. As the velocity of the swimmer increases the learning process becomes faster. {\bf C)} Contrast between the best and the next best action $\Delta G_k$ after learning according to equation \ref{eq:contrast} for two different Peclet numbers ($Pe=40$ and $Pe=120$). The results demonstrate that the swimmer is less decided in noisier environment. {\bf D)} Policy obtained from learning processes at low and high velocity ($Pe =40$ and $Pe=120$). The orange boxes mark areas, where the learning results in less risky actions for low Peclet numbers. {\bf E)} Trajectories of a swimmer obtained after reinforcement learning in a gridworld with virtual obstacles defined by penalties (red squares) and a goal state next to the obstacle (gray square). {\bf F)} Learning progress (sum of all Q-matrix values) for one and two swimmers in the gridworld ($Pe=80$). The two swimmers share the same Q-matrix, which they update at each step.} 
	\label{fig3}
\end{figure*}

The application of the Q-learning algorithm to our artificial microswimmer introduces a correlation of the actions in time\cite{Sutton1998}. This correlation and the restricted values of $\alpha$ and $\gamma$ cause the convergence of values in the Q-matrix (equation \ref{eq1}) and its sum. Figure \ref{fig2} summarizes the learning process for a single swimmer in the gridworld for a Peclet number of $Pe=80$. The learning is visible in the decrease of the number of steps required to reach the target (Fig. \ref{fig2}A inset) and the persistence of the trajectories (Fig. \ref{fig2}B). The random initial values in the Q-matrix cause a random selection of actions at the beginning of the learning process and make the swimmer to move in loops until it eventually reaches the boundary or the goal. Within the initial episodes of the learning process the swimmer needs about 90 transitions to reach the goal if starting at the lower left state. The reward of the goal and the penalty of the boundary propagates to all states of the gridworld and correlates the actions in different states so that the trajectories run increasingly straight to the goal. The policy after more than 300 learning episodes (5500 steps and nearly converged Q-matrix sum) leads to a directed motion towards the goal now requiring only about 7 transitions if starting from the lower left state. The directed motion towards the goal is independent of which state is used as the starting point (Fig. 2C, and video 2 in the SI). The corresponding policy at this stage of the learning reflects this effective "flow" of the swimmer towards the target state and is clearly different from the initial policy after the initialization of the Q-matrix (Fig. \ref{fig2}D). 

The influence of Brownian motion can now be considered in two different situations. First, the learned policy for a given Peclet number can be provided to other swimmers with higher or lower Peclet number. 
Figure \ref{fig3}A displays the density maps of 50 trajectories for swimmers with different Peclet numbers ($Pe = 53$, $Pe = 91$, $Pe = 125$, $Pe = 135$), all moving with the policy obtained for $Pe=80$ (Fig. \ref{fig2}D). All trajectories start in the lower left corner and reach the goal at the top right corner. The black line shows an example of a trajectory for each velocity and illustrates that the motion is becoming more persistent as the Peclet number is increased. The average path length decreases from $46\, {\rm \mu m}$ at $Pe = 53$ to $33\, {\rm \mu m}$ at $Pe = 135$ and the average time to reach the goal decreases from $18\, {\rm s}$ at $Pe = 53$ to only $6\, {\rm s}$ at $Pe = 135$. Therefore, the learned policy seems to be a robust solution also for swimmers with a different noise level.
 Second, the level of noise may also influence the learning process itself, as a more deterministic motion where the applied action leads to the intended state change can increase the learning speed as displayed in Fig. \ref{fig3}B. At $Pe=150$ the system is almost converged after 1000 learning steps, while it requires more than 5000 steps at $Pe=87$. At very low Peclet numbers the learning process will be very time consuming. The nearly converged Q-matrix at different Peclet numbers reveals that the contrast between the best and the next best actions as determined by 
\begin{equation}\label{eq:contrast}
\Delta G_k =\frac{1}{N}\sum_{i=1}^{N} Q(s_i,a_i^{b})-Q(s_i,a_i^{k}) 
\end{equation}
is diminished for increasing noise in the motion (Fig. \ref{fig3}C). Here $N$ is the number of states and $k$ is the index in the i-th vector that holds the Q values for the state $s_i$. The swimmer is thus less certain which action to take with decreasing Peclet number. This also leads to the fact that the swimmer is trying to find a less "risky" policy for low Peclet numbers. Figure \ref{fig3}D indicates that especially the actions at the boundary (orange boxes) are pointing more inwards at low Peclet number as compared to the high Peclet number solution ($Pe=40$ and $Pe=120$ in Fig. \ref{fig3}D). The swimmer therefore avoids regions close to the boundary if the system is more noisy, which is very similar to the cliff-walking problem in RL. 
Coming back to the first discussed reuse of policies for other swimmers, this result also means that using a policy learned at high Peclet number with a swimmer moving at lower Peclet number gets more risky for the swimmer especially at the boundary (Fig. \ref{fig3}A). Less deterministic swimmers thus require to learn a different policy in order to reach the goal.
The presented control method employed for the RL of microswimmers can also be extended to contain virtual obstacles. These are introduced by strong penalties for certain states. In contrast to real obstacles, which reduce the number of accessible states in the sample space, virtual obstacles can be occupied by the swimmer at the cost of a penalty. Figure \ref{fig3}E displays examples where virtual obstacles have been introduced with a goal state right next to these virtual walls (Fig. \ref{fig3}E and S.I.). The learning process is then conducted in the same way as previously described and the system learns to avoid the obstacle to reach the goal. Figure \ref{fig3}E also depicts recorded example trajectories for the swimmers after the Q-matrix is sufficiently converged (see video 3 and S.I.). 

So far we have considered the learning process for a single microswimmer. More complex situations may involve the emergence of collective behavior, where the motion of multiple agents is concerned. Different levels of collectivity and coorperative learning may be addressed \cite{Sheng2015, Schutter2006}. A true collective learning is carried out when the swimmer is taking an action to maximize the reward of the collective, not only its individual one. Swimmers may also learn to act as a collective when the rewards are given if an agent behaves like others in an ensemble \cite{Ried:it}. This mimics the process of developing swarming behavior implicated by the Vicsek model for example \cite{Vicsek}. Our control mechanism is capable of addressing multiple swimmers separately such that they may also cooperatively explore the environment. Instead of a true collective strategy, we are considering a low density of swimmers (number of swimmers $\ll $ number of states). The swimmers in the gridworld will share their information during the learning process, meaning that all swimmers draw their actions from the same Q-matrix which they also update cooperatively. The swimmers are exploring the same gridworld in different spatial regions and thus a speed-up of the learning is expected. 
Figure \ref{fig3}F shows a comparison of the sum of Q-matrix elements as a function of time for the single- and the two-swimmer case exploring a gridworld with the upper right corner as a goal state (both processes at the same $Pe = 80$). The two-particle system converges clearly faster demonstrating the benefit of sharing information between individuals of a group in a learning process. 

The integration of reinforcement learning into the world of microswimmers we have presented is a hybrid solution, where the agent is a real-world object in a real-world environment, but its "brain" and sensing capabilities are externalized to a computer and a microscopy setup. Already with this hybrid solution one obtains a simple scalable model system, where strategies in a noisy environment with virtual obstacles or collective learning can be explored. As compared to a complete computer simulation, our system contains a non-ideal control, meaning that liquid flows, imperfections of the swimmers or sample container, hydrodynamic interactions or other uncontrolled parameters naturally influence the learning process. In that sense even the inverse problem of using the learned strategy to reveal the details of these uncontrolled influences may be tackled as a new form of environmental sensing. While the implementation of signaling and feedback by physical or chemical processes into a single artificial microswimmer is still a distant goal, the current hybrid solution opens a whole branch of new possibilities for understanding adaptive behavior of single microswimmers in noisy environments and the emergence of collective behavior of large ensembles of active systems. 

\section*{Methods}

{\bf{Sample.}} Samples consist of commercially available  gold nanoparticle coated melamine resin particles of a diameter of 2.19 $\rm \mu m$ (microParticles GmbH, Berlin, Germany). The gold nanoparticles are covering about 30 \% of the surface and are between 8 and 30 nm in diameter. Microscopy glass cover slides have been dipped into a 5 \% Pluronic F127 solution and rinsed with deionized water and dried with nitrogen. The Pluronic F127 coating prevents sticking of the particles on the glass cover slides. Two microliters of particle suspension are placed on the cover slides to spread about an area of 1 cm $\times$ 1 cm forming a 3 $\rm \mu m$ thin water film. The edges of the sample have been sealed with silicone oil (PDMS) to prevent water evaporation.
\\
\\
{\bf{Experimental Setup.}} Samples have been investigated in a custom-built inverted dark-field microscopy setup. The sample region is illuminated by a dark-field oil-immersion condenser (Olympus, NA 1.2). The scattered light is collected by an oil-immersion objective lens (Olympus 100x, NA 1.35-0.6) with the numerical aperture set to 0.6 and imaged to an Andor iXon emCCD camera. A $\lambda=532\, {\rm nm}$ laser is focused by the imaging objective into the sample plane to serve as a heating laser for the swimmers. Its position in the sample plane is steered by an acousto-optic deflector (AOD, AA Opto-Electronic) together with a 4-f system (two $f=20$ cm lenses). The AOD is controlled by an AdWin realtime board (AdWin Gold, J\"ager Messtechnik), which is exchanging data with a custom LabView program. A region of interest (ROI) of 512 x 512 pixels (30 $\rm \mu m $ x 30 $\rm \mu m$) is utilized for the real-time imaging, analysis and recording of the particles, with an exposure time of 180 ms. More details can be found in the supplementary information.


\section*{Acknowledgements}
The authors acknowledge financial support by  the DFG priority program 1726 "Microswimmers" throughout the project CI 33/16-2. S.M. acknowledges fruitful discussions and comments of Emilio, Jose and Ricardo Faro.

\end{document}